\documentclass[aps,prl,reprint,groupedaddress]{revtex4-1}

\usepackage{amsmath}
\usepackage{graphicx}
\usepackage{epstopdf}

\begin{document}

\title{Custom-tailored spatial mode sorting by controlled random scattering}

\author{Robert Fickler$^{1}$}
\author{Manit Ginoya$^{1}$}
\author{Robert W. Boyd$^{1,2}$}
\affiliation{$^1$Department of Physics, University of Ottawa, Ottawa, Canada
\\
$^2$Institute of Optics, University of Rochester, Rochester, USA
}

\begin{abstract} 
\textbf{The need to increase data transfer rates constitutes a key challenge in modern information-driven societies. Taking advantage of the transverse spatial modes of light to encode more information is a promising avenue for both classical and quantum photonics. However, to ease access to the encoded information, it is essential to be able to sort spatial modes into different output channels. Here, we introduce a novel way to customize the sorting of arbitrary spatial light modes. Our method relies on the high degree of control over random scattering processes by preshaping of the phase structure of the incident light. We demonstrate experimentally that various sets of modes, irrespective of their specific modal structure, can be transformed to any output channel arrangement. Thus, our method enables full access to all of the information encoded in the transverse structure of the field, for example, azimuthal and radial modes. We also demonstrate that coherence is retained in this complex mode transformation, which opens up applications in quantum and classical information science.}
\end{abstract}

\date{\today}

\maketitle

Transverse spatial light modes have attracted a much attention because of their special properties and the subsequent broad range of applications \cite{andrews2012angular,rubinsztein2016roadmap}. One particularly promising approach uses spatial modes of light to increase data rates in optical communication schemes. By not only harnessing the polarization or frequency of light but also the transverse spatial degree of freedom, a dramatic improvement in the multiplexing of information can be achieved \cite{wang2012terabit,willner2015optical}. In quantum physics spatial light modes are successfully utilized as laboratory realizations of high-dimensional quantum states \cite{mair2001entanglement,krenn2014generation}, which are advantageous for quantum secure communication schemes \cite{mirhosseini2015high}, quantum simulation tasks \cite{cardano2015quantum} and foundational investigations \cite{malik2016multi}.
However, to fully take advantage of the potential that spatial light modes offer, technologies to manipulate and measure them are essential. Although mature technologies to generate complex spatial modes are available today \cite{forbes2016creation}, complex transformations of such modes are still rare and challenging to implement \cite{morizur2010programmable,leach2002measuring,schlederer2016cyclic}. The transformation from a given spatial mode to a specific position in a transverse plane, that is, mode sorting or demultiplexing, is an especially interesting transformation in quantum \cite{mirhosseini2015high,fickler2014interface,malik2014direct} and classical information schemes \cite{willner2015optical,labroille2014efficient}. A specific example, namely a sorter for modes that differ by their azimuthal structure, has recently been established \cite{berkhout2010efficient,mirhosseini2013efficient} and successfully implemented \cite{mirhosseini2015high,fickler2014interface,malik2014direct,willner2015optical}.

Here we show that sorting of this type can be implemented through the control of strong scattering processes, a topic that has recently attracted much attention \cite{mosk2012controlling}. Such control can allow one to realize complex modulation tasks that have thus far been impossible. Earlier experiments presage the enormous potential that the control over random scattering processes can offer, by e.g. showing enhanced transmission and focusing through opaque, scattering media in the spatial \cite{vellekoop2010exploiting,bertolotti2012non,katz2012looking} and temporal domains \cite{katz2011focusing}. More importantly, it was also demonstrated that controlled scattering can be used to realize a programmable beam splitter \cite{huisman2015programmable} or to investigate complex quantum walks \cite{defienne2016two}.

\subsection{The custom-tailored sorting scheme}
\begin{figure}[b]
\centering  \includegraphics[width=0.44\textwidth ]{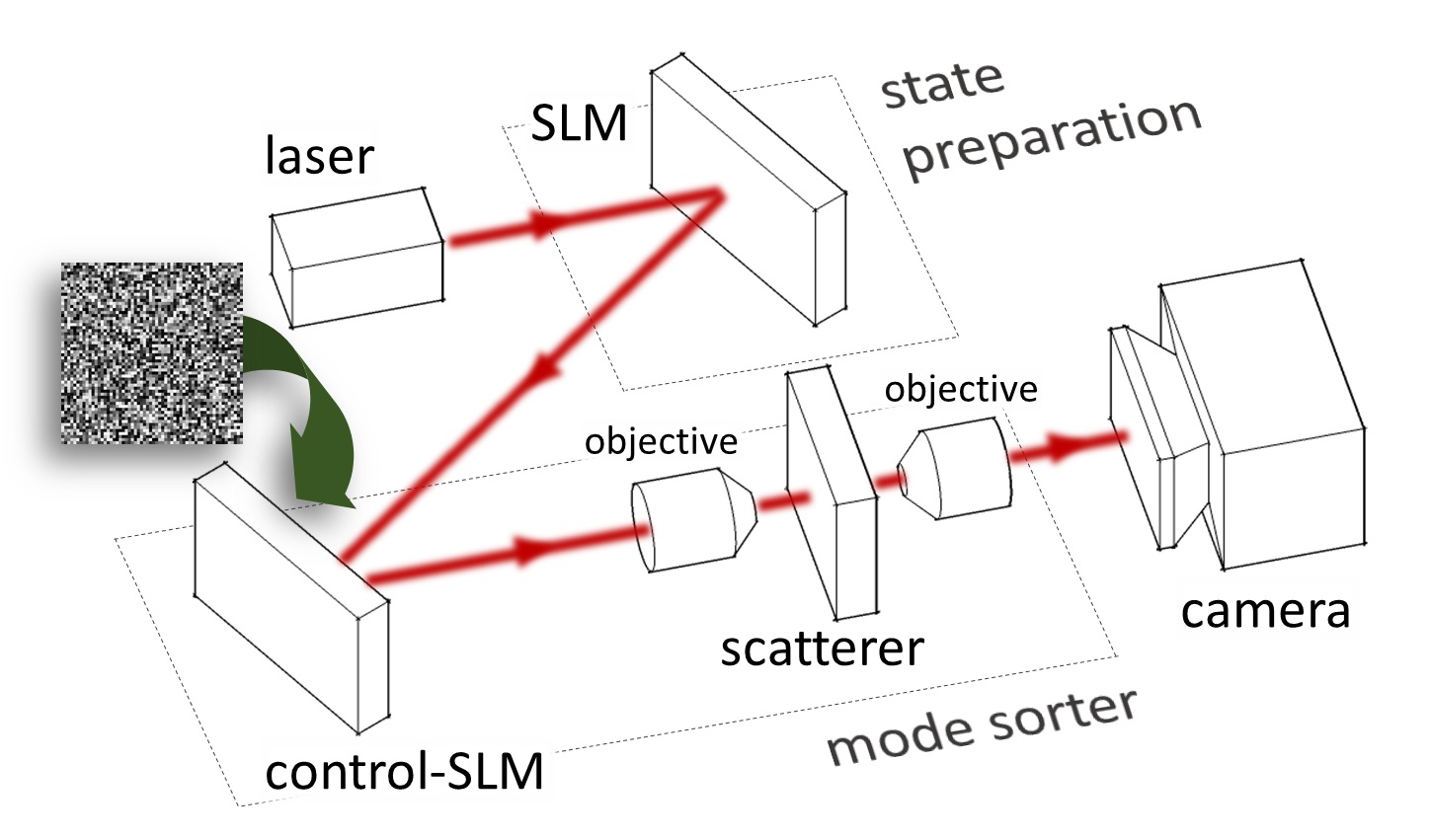}
\caption{Sketch of the experimental setup for custom-tailored spatial mode sorting. Spatial light modes are generated by modulating the transverse spatial structure of a laser beam by means of a spatial light modulator (SLM). The laser light is then redirected onto a control-SLM and focused into a strongly scattering medium (TiO$_2$ powder). The appropriate phase modulation structure of the control-SLM (inset left side, gray value stands for phase modulation depth up to 2$\pi$) is found by use of a genetic algorithm and the feedback signal from a CMOS camera. For more details see main text. \label{fig1}}
\end{figure}

In our experimental setup (shown in Fig. \ref{fig1}) we first generate spatial light modes by diffracting a laser beam of a computer-generated hologram displayed on a phase-only spatial light modulator (SLM). With this technique, we can imprint any desired phase structure onto beams in the first diffraction order and thus can prepare the desired light modes and their superpositions \cite{forbes2016creation}. When the beams have a radial profile (see examples in Fig. \ref{fig5}a), we additionally modify the radial structure the amplitude of the beam by appropriate shaping of the grating efficiency \cite{bolduc2013exact}. After generating the modes we send them onto another SLM, the control-SLM, which modulates only the phase and thereby controls the scattering process. We then image the plane of the control-SLM onto the input aperture of a microscope objective (60x), which focuses the light into the scattering sample. The scatterer consists of a layer of TiO$_2$ powder of $11.3\pm 0.9$ $\mu$m thickness leading to multiple scattering processes (estimated mean free path on the order of 0.5 $\mu$m - 1 $\mu$m \cite{kop1997observation}), which causes strong coupling between the different input channels and thus complex modulations. The scattered light is collected by a second microscope objective (20x) and recorded by a CMOS camera. Following earlier experiments \cite{mosk2012controlling,vellekoop2010exploiting,bertolotti2012non,katz2012looking,katz2011focusing,huisman2015programmable,defienne2016two}, we use a feedback signal, in our case the recorded image, and a genetic algorithm \cite{conkey2012genetic} to optimize the modulation of the control-SLM for the best sorting of the input light modes.

An experimental optimization run contains the following steps: We first generate 30 different random patterns (population), each consisting of 60x60 squares which in turn are built of 10x10 pixels (see example in Fig. \ref{fig1} inset on the left). Each square modulates the phase between 0 and 2$\pi$ individually in the course of the optimization process. Moreover, a blazed grating with a grating period of 15 pixels is overlaid over the whole phase modulation such that only the controllable first diffraction order is focused onto the scatterer. We then display all patterns on the SLM to evaluate their sorting abilities (fitness). The evaluation of each fitness is done by sending all modes sequentially through the sorter, i.e. the combination of control-SLM displaying the pattern and the scatterer, and recording the scattered intensities at all pre-defined sorting locations (each approx. 21 $\mu$m x 21 $\mu$m or 4x4 pixels in size) with the camera. The algorithm then calculates the fitness for the control pattern as the sum over all intensities found at the different locations to which the modes should be sorted. Additionally, it subtracts the undesirable overlap to the other sorting channels, i.e. intensities found at the sorting locations for the other modes. With this subtraction we optimize not only for maximal sorting efficiency but also for minimal overlap between the sorted modes. After this initial evaluation the algorithm optimizes autonomously for best sorting in a typical genetic-algorithm manner, i.e. by combining the best patterns, adding slight mutations, evaluating the fitness, and replacing the patterns if the newly generated ones are superior. In general, we find that after approximately 5000 iterations the sorting ability saturates leading to an optimal sorting. However, whenever 5 or more modes are sorted up to 15000 iterations were required. The speed of this optimization was mainly limited by the refresh rate of the SLMs (max. 50 Hz) and the read out of the camera. The latter required an averaging over many snapshots to compensate for an intensity beating induced by the flickering of two sequentially used SLMs.

\subsection{Sorting of modes with different azimuthal structure}

\begin{figure}[tb]
\centering  \includegraphics[width=0.43\textwidth ]{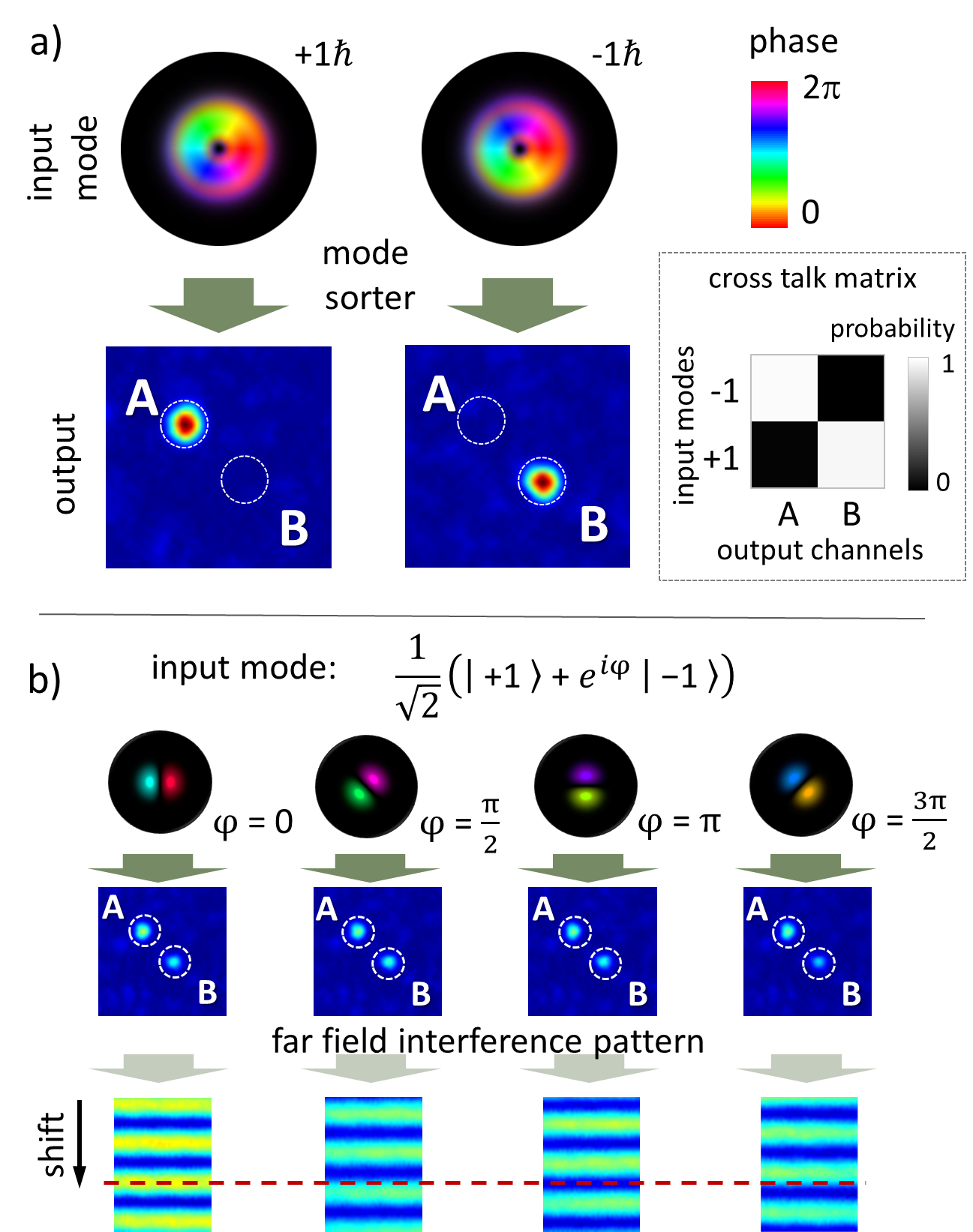}%
\caption{Sorting of spatial modes carrying $\pm$1 unit of OAM. a) Upper part: Either of two OAM modes is sent into the sorter. The color depicts their phase structure while the brightness shows the intensity distribution. Lower part: Experimentally recorded intensities after transmission of either mode through the setup. We find OAM modes with +1$\hbar$ and -1$\hbar$ of OAM to be sent to the two predefined output channels A and B, respectively (false color recording). Inset: Cross-talk matrix of the sorter from which a sorting ability of 97.7$\pm$0.6\% can be deduced. b) Coherence test of the mode-sorting process. If equally weighted superpositions are sent through the sorter, both output channels A and B show equal intensities for different phases $\varphi$. After propagation to the far field, an interference pattern ($\sim$88\% visibility) is found, which shifts laterally depending on the input phases of the OAM superpositions. \label{fig2}}
\end{figure}

As a first example, we generate and sort up to 7 orthogonal modes each having a specific helical phase front $e^{il\phi}$, with $\phi$ being the angle and $l$ an integer value between -3 and 3 describing the orbital angular momentum (OAM) quantum number \cite{mair2001entanglement}. To quantify the achieved sorting we evaluate the sorting probability $P_n$ for each mode out of the set of $m$ different modes:  $P_n=\frac{I_n}{\Sigma_m I_m}$. Here $I_n$ stands for the intensity at the $n^{th}$ sorting positions, which is measured as the sum over all intensities found in the target region in the recorded camera image. We define the average over all sorting probabilities as the sorting ability, which is used to quantify the sorting process.
We start by sorting only two modes ($\pm$1 OAM), and we obtain a sorting ability of 97.7$\pm$0.6\% into the predefined output channels after 5000 iterations, i.e. 20 minutes of optimization time, (see Fig. \ref{fig2}a). We then test the coherence of the process by sending equally weighted coherent superpositions of the two modes through the sorter. Irrespective of the phase between the two superimposed modes, we find equal intensities in the two output channels,  (see Fig. \ref{fig2}b). To investigate if the phases are conserved, we place two pinholes at the output channel locations and interfere the sorted light with the help of a lens placed behind the pinholes. The resulting interference pattern in the focal plane, i.e. far-field, shows a visibility of approximately 88\%. The slight deviations from a perfect visibility stem from residual stray light and imperfect mode sorting. More importantly, a change of the phase between the two OAM modes before sorting, which gives rise to a different rotation angle of the superposition structure, leads to a clear shift of the interference fringes (see Fig. \ref{fig2}b). Thus the process preserves coherence among the sorted modes, an essential feature for quantum or classical information processing. 

\begin{figure}[tb]
\centering  \includegraphics[width=0.49\textwidth ]{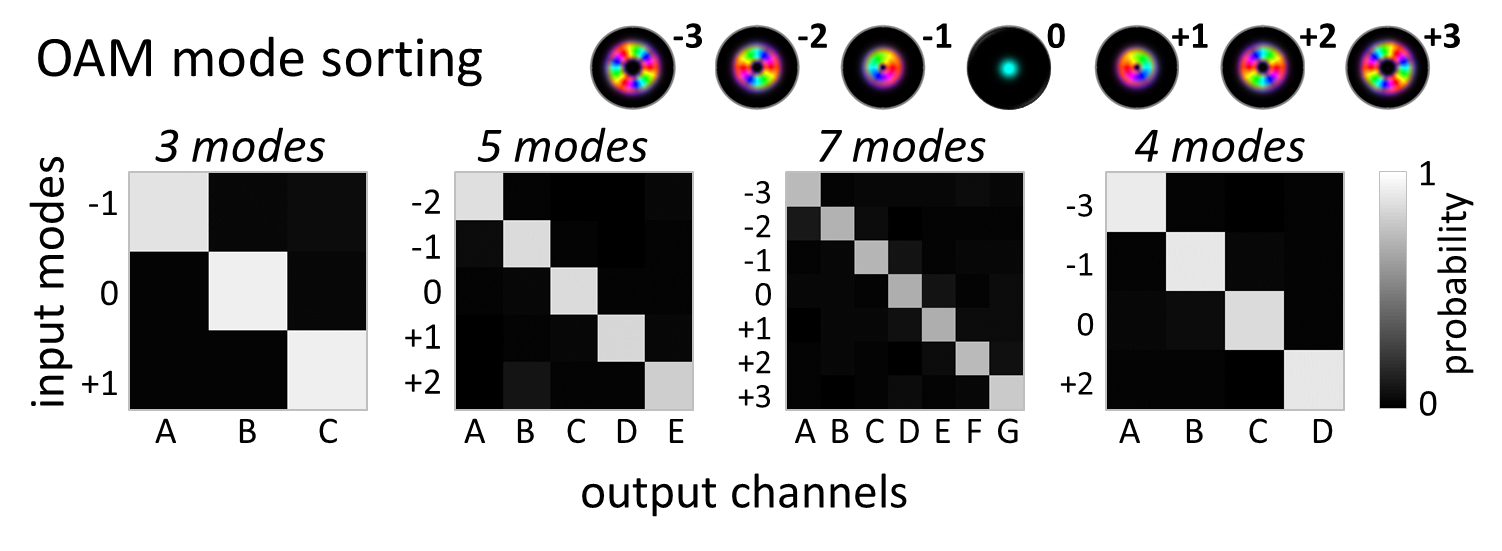}%
\caption{Sorting of up to seven OAM modes carrying up to $\pm$3 quanta of OAM. The sorter can be programmed to sort 3, 5 and 7 neighbouring and 4 randomly chosen OAM modes. All measurements demonstrate very good sorting abilities of 94$\pm$1\%, 85$\pm$3\%, 72$\pm$4\% and 90$\pm$3\%, respectively, which is nicely visualized by the high probabilities along the diagonal axis of the cross-talk matrices. \label{fig3}}
\end{figure}

Moreover, we investigated the ability to sort a greater number of modes and to analyze the stability of our device. We are able to sort the three, five, and seven lowest-order OAM modes as well as 4 randomly chosen modes with a 2 sorting ability of 94$\pm$1\%, 85$\pm$3\%, 72$\pm$4\% and 90$\pm$3\%, respectively (see Fig. \ref{fig3}). Note that the decreased sorting ability for a larger number of modes originates from the increased number of output channels, which all have a small but non-zero overlap to the other modes. If the sorting ability of only two modes out of the seven modes is evaluated, we find similar results as before of around 95\% sorting ability. Additionally, sorting a larger number of modes requires a larger optimization time of around 45 min, 90 min, 240 min and 60 min, respectively. To investigate the long-time stability of this sorting scheme, we measure the sorting ability for 5 modes over the course of 24 hours without re-optimization, and we find a small decrease from 85\% to 84\%. This result demonstrates the very high stability of our scheme. It also opens the possibility to implement different sorting configurations (for different modes or different output channel geometries, see below) in one setup between which one can switch in real-time with the speed of the refresh rate of the control-SLM.

\subsection{Flexibility of output channel geometry}

\begin{figure}[b]
\centering  \includegraphics[width=0.49\textwidth ]{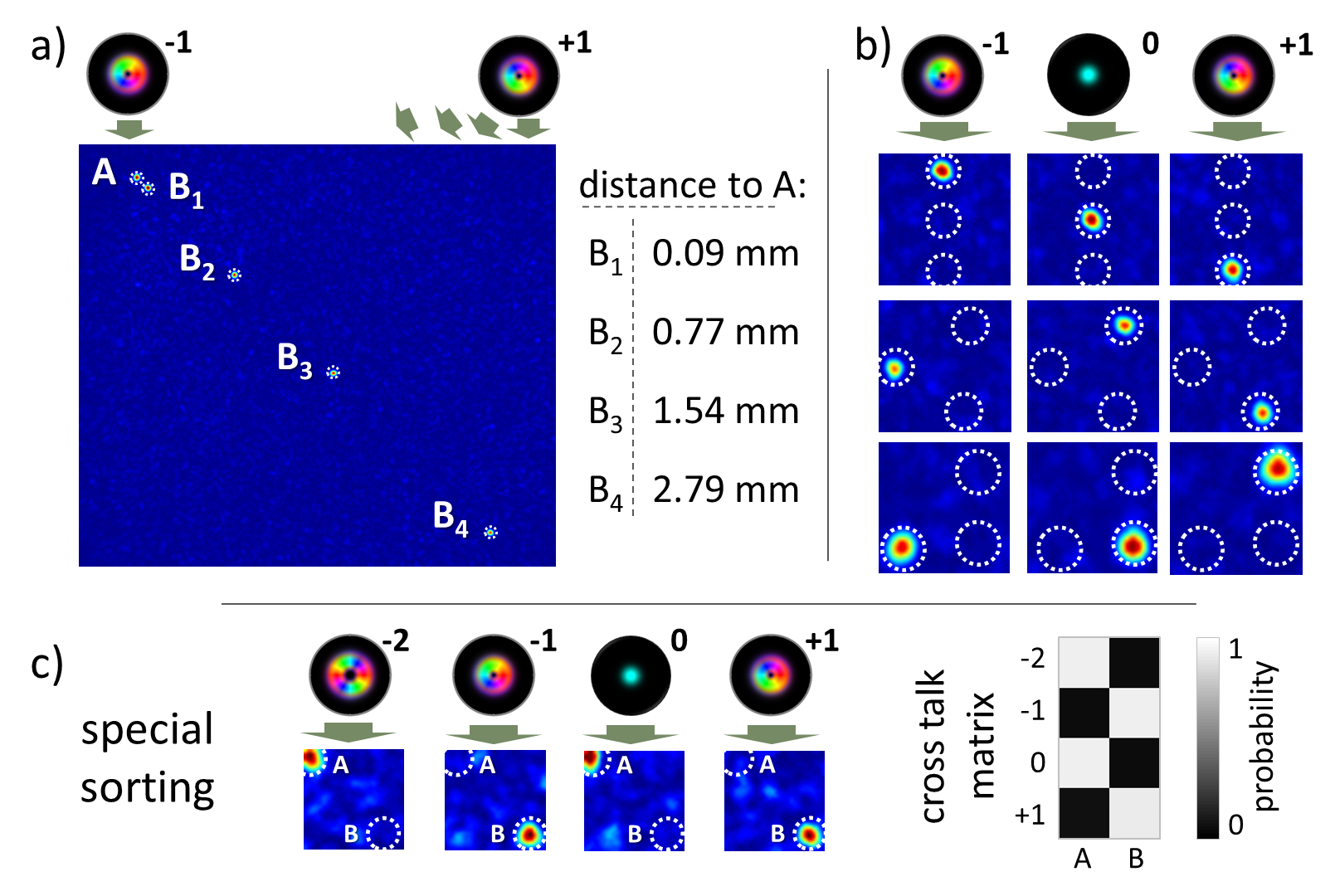}%
\caption{Demonstration of sorting to different output-channel arrangements. (a) The distance between the two output locations A and B can be adjusted from 88 $\mu$m to 2.8 mm without a detectable reduction of sorting ability (97\% -98\%). Note that the four results were recorded in separate runs but are embedded in one image here to illustrate the distances graphically. b) The arrangement of the output channels can be arbitrarily chosen with similar sorting abilities of approximately 93\%. Examples of vertical, triangular, and corner-shape sorting of three modes are shown. c) Different input modes can be sorted to the same output position, e.g. -2 and 0 to A, -1 and +1 to B, with no reduction on the sorting ability (see cross-talk matrix). Color coding of modes as in Fig. \ref{fig2}a. \label{fig4}}
\end{figure}

In a second series of measurements, we test the flexibility of the output channel geometry, that is, the distance between output channels and their arrangement. We start again by sorting two modes ($\pm$1 OAM) and optimize for different distances between the sorting locations (see Fig. \ref{fig4}a). With almost the same sorting ability (around 97\% -98\%) we can change the distance of the output ports from 88 $\mu$m up to 2.8 mm. The minimum distance is limited by the achievable spot size after scattering and the technical limitation of the feedback, that is, the pixel size of the camera, while the collection aperture of the second microscope objective restricts the maximum distance between the spots. We then test different output channel arrangements for three modes (0$\hbar$, $\pm$1$\hbar$ OAM). All tested arrangements, vertical, triangular and corner-shape geometries (see Fig. \ref{fig4}b), show similar sorting abilities (around 93\%), thus the sorter can be adjusted to any specific experimental requirements. Interestingly, different modes can even be sorted into the same location (see Fig. \ref{fig4}c). Although outside the scope of this demonstration, it will be interesting to investigate whether this sorting scheme requires scattering losses that lead to a reduced overall efficiency. In our case only up to 1\% of the input intensity was sorted into the desired output channels. The use of multimode fibers, which have also been used to achieve complex scattering \cite{defienne2016two}, should reduce the losses substantially, however, with a possible expense of losing stability.

\subsection{Sorting of radial modes and Hermite Gauss modes}

\begin{figure}[b]
\centering  \includegraphics[width=0.43\textwidth ]{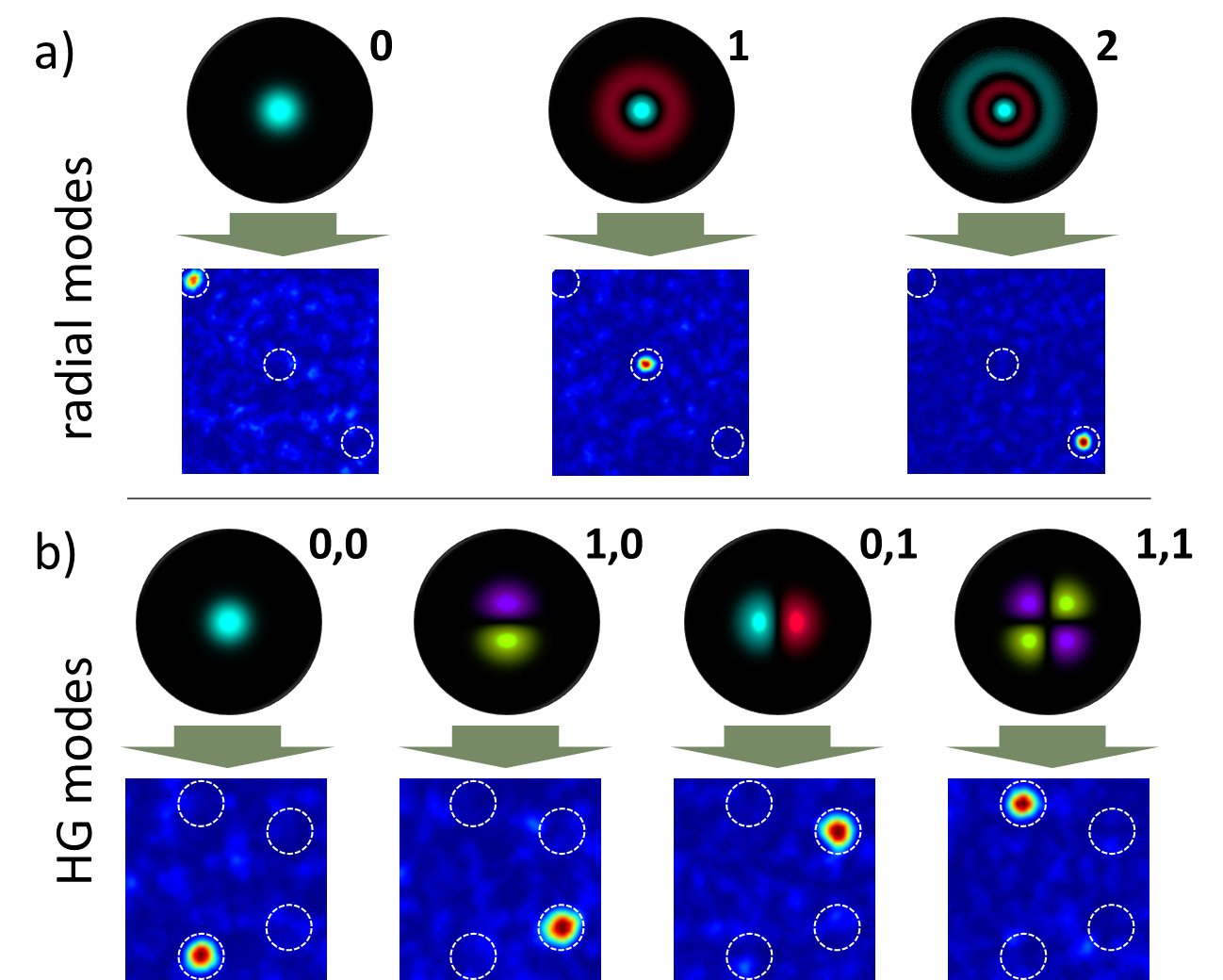}%
\caption{Sorting of different types of light modes. a) Radial modes that differ only by their radial structure are sorted to different spatial locations and high sorting ability of 92$\pm$2\%. b) The four lowest-order Hermite-Gauss (HG) modes are sorted with a sorting ability of 85$\pm$3\%. Color coding for all modes as in Fig. \ref{fig2}a. \label{fig5}}
\end{figure}

So far we sorted different twisted light modes, that is, modes that differ by their azimuthal phase structure, for which, as mentioned earlier, a highly efficient sorter already exists \cite{berkhout2010efficient,mirhosseini2013efficient}. However, our technique is not limited to modes with different azimuthal phases but works for all orthogonal spatial modes of light. To demonstrate this feature, we investigate modes that have no helical phase dependence and differ only by their radial structure (radial $\pi$-phase steps), i.e. radial modes. Together with the azimuthal degree of freedom, they complete the Laguerre-Gauss (LG) mode basis, which is a preferred mode set in a system with cylindrical symmetry \cite{zhao2015capacity,kahn2016twist}. To generate the three lowest-order radial modes, we use an amplitude- and phase-modulation technique \cite{bolduc2013exact} before sending them through the scatterer. Similarly as before, the genetic algorithm is capable of determining a phase pattern for the control-SLM that leads to high sorting ability of 92$\pm$2\% (see Fig. \ref{fig5}a). Thus, we show that our technique is capable of decomposing light fields into a complete set of modes, e.g., its LG modes. As a last test, we generate and sort four modes of a different mode family and symmetry: Hermite-Gauss modes, which differ by vertical or horizontal p-phase steps (see Fig. \ref{fig5}b). The sorting ability of 85$\pm$3\% is comparable to the results for four twisted-light modes and demonstrates that not only LG modes but also different mode families can be sorted.

In summary, we have shown that it is possible to realize custom-tailored mode sorting by means of manipulating strong scattering processes by placing an appropriate phase structure onto the input light field. We used the high degree of control over strongly scattering media to custom-tailor the sorting process with respect to the number and type of input spatial light modes as well as to the output channel arrangement. Our results might be beneficially applied in the various different fields of classical and quantum optics using structured light \cite{andrews2012angular,rubinsztein2016roadmap}. By having access to information encoded in the radial and azimuthal degrees of freedom, the full potential of transverse light modes to increase data rates can be exploited \cite{wang2012terabit,willner2015optical,zhao2015capacity,kahn2016twist}. The demonstrated coherence also paves the path to applications in quantum information science, where photons residing in a high-dimensional state space might also benefit from such transformations. For such applications, it will be important to increase the overall efficiency of the process, e.g. by using multi-mode fibers as complex scattering media \cite{defienne2016two}. Other fascinating open questions are the relation among transformation abilities, precision of the control, and influence of losses.

\begin{acknowledgments}
\subsection{acknowledgments}
We thank J. Upham and G. Poirier for helping with the investigation of the scattering samples, J. Horton for his help at an early stage of the experiment and E. Giese, R. Lapkiewicz and M. Krenn for helpful discussions. R.F. acknowledges the support of the Banting postdoctoral fellowship of the Natural Sciences and Engineering Research Council of Canada (NSERC). R.F., M.G. and R.W.B. gratefully acknowledge the support of the Canada Excellence Research Chairs Program (CERC).
\end{acknowledgments}

\bibliography{BIB_Fickler_Scatterer}{}

\begin{thebibliography}{30}%
\makeatletter
\providecommand \@ifxundefined [1]{%
 \@ifx{#1\undefined}
}%
\providecommand \@ifnum [1]{%
 \ifnum #1\expandafter \@firstoftwo
 \else \expandafter \@secondoftwo
 \fi
}%
\providecommand \@ifx [1]{%
 \ifx #1\expandafter \@firstoftwo
 \else \expandafter \@secondoftwo
 \fi
}%
\providecommand \natexlab [1]{#1}%
\providecommand \enquote  [1]{``#1''}%
\providecommand \bibnamefont  [1]{#1}%
\providecommand \bibfnamefont [1]{#1}%
\providecommand \citenamefont [1]{#1}%
\providecommand \href@noop [0]{\@secondoftwo}%
\providecommand \href [0]{\begingroup \@sanitize@url \@href}%
\providecommand \@href[1]{\@@startlink{#1}\@@href}%
\providecommand \@@href[1]{\endgroup#1\@@endlink}%
\providecommand \@sanitize@url [0]{\catcode `\\12\catcode `\$12\catcode
  `\&12\catcode `\#12\catcode `\^12\catcode `\_12\catcode `\%12\relax}%
\providecommand \@@startlink[1]{}%
\providecommand \@@endlink[0]{}%
\providecommand \url  [0]{\begingroup\@sanitize@url \@url }%
\providecommand \@url [1]{\endgroup\@href {#1}{\urlprefix }}%
\providecommand \urlprefix  [0]{URL }%
\providecommand \Eprint [0]{\href }%
\providecommand \doibase [0]{http://dx.doi.org/}%
\providecommand \selectlanguage [0]{\@gobble}%
\providecommand \bibinfo  [0]{\@secondoftwo}%
\providecommand \bibfield  [0]{\@secondoftwo}%
\providecommand \translation [1]{[#1]}%
\providecommand \BibitemOpen [0]{}%
\providecommand \bibitemStop [0]{}%
\providecommand \bibitemNoStop [0]{.\EOS\space}%
\providecommand \EOS [0]{\spacefactor3000\relax}%
\providecommand \BibitemShut  [1]{\csname bibitem#1\endcsname}%
\let\auto@bib@innerbib\@empty
\bibitem [{\citenamefont {Andrews}\ and\ \citenamefont
  {Babiker}(2012)}]{andrews2012angular}%
  \BibitemOpen
  \bibfield  {author} {\bibinfo {author} {\bibfnamefont {D.~L.}\ \bibnamefont
  {Andrews}}\ and\ \bibinfo {author} {\bibfnamefont {M.}~\bibnamefont
  {Babiker}},\ }\href@noop {} {\emph {\bibinfo {title} {The angular momentum of
  light}}}\ (\bibinfo  {publisher} {Cambridge University Press},\ \bibinfo
  {year} {2012})\BibitemShut {NoStop}%
\bibitem [{\citenamefont {Rubinsztein-Dunlop}\ \emph
  {et~al.}(2016)\citenamefont {Rubinsztein-Dunlop}, \citenamefont {Forbes},
  \citenamefont {Berry}, \citenamefont {Dennis}, \citenamefont {Andrews},
  \citenamefont {Mansuripur}, \citenamefont {Denz}, \citenamefont {Alpmann},
  \citenamefont {Banzer}, \citenamefont {Bauer} \emph
  {et~al.}}]{rubinsztein2016roadmap}%
  \BibitemOpen
  \bibfield  {author} {\bibinfo {author} {\bibfnamefont {H.}~\bibnamefont
  {Rubinsztein-Dunlop}}, \bibinfo {author} {\bibfnamefont {A.}~\bibnamefont
  {Forbes}}, \bibinfo {author} {\bibfnamefont {M.}~\bibnamefont {Berry}},
  \bibinfo {author} {\bibfnamefont {M.}~\bibnamefont {Dennis}}, \bibinfo
  {author} {\bibfnamefont {D.~L.}\ \bibnamefont {Andrews}}, \bibinfo {author}
  {\bibfnamefont {M.}~\bibnamefont {Mansuripur}}, \bibinfo {author}
  {\bibfnamefont {C.}~\bibnamefont {Denz}}, \bibinfo {author} {\bibfnamefont
  {C.}~\bibnamefont {Alpmann}}, \bibinfo {author} {\bibfnamefont
  {P.}~\bibnamefont {Banzer}}, \bibinfo {author} {\bibfnamefont
  {T.}~\bibnamefont {Bauer}},  \emph {et~al.},\ }\href@noop {} {\bibfield
  {journal} {\bibinfo  {journal} {Journal of Optics}\ }\textbf {\bibinfo
  {volume} {19}},\ \bibinfo {pages} {013001} (\bibinfo {year}
  {2016})}\BibitemShut {NoStop}%
\bibitem [{\citenamefont {Wang}\ \emph {et~al.}(2012)\citenamefont {Wang},
  \citenamefont {Yang}, \citenamefont {Fazal}, \citenamefont {Ahmed},
  \citenamefont {Yan}, \citenamefont {Huang}, \citenamefont {Ren},
  \citenamefont {Yue}, \citenamefont {Dolinar}, \citenamefont {Tur} \emph
  {et~al.}}]{wang2012terabit}%
  \BibitemOpen
  \bibfield  {author} {\bibinfo {author} {\bibfnamefont {J.}~\bibnamefont
  {Wang}}, \bibinfo {author} {\bibfnamefont {J.-Y.}\ \bibnamefont {Yang}},
  \bibinfo {author} {\bibfnamefont {I.~M.}\ \bibnamefont {Fazal}}, \bibinfo
  {author} {\bibfnamefont {N.}~\bibnamefont {Ahmed}}, \bibinfo {author}
  {\bibfnamefont {Y.}~\bibnamefont {Yan}}, \bibinfo {author} {\bibfnamefont
  {H.}~\bibnamefont {Huang}}, \bibinfo {author} {\bibfnamefont
  {Y.}~\bibnamefont {Ren}}, \bibinfo {author} {\bibfnamefont {Y.}~\bibnamefont
  {Yue}}, \bibinfo {author} {\bibfnamefont {S.}~\bibnamefont {Dolinar}},
  \bibinfo {author} {\bibfnamefont {M.}~\bibnamefont {Tur}},  \emph {et~al.},\
  }\href@noop {} {\bibfield  {journal} {\bibinfo  {journal} {Nature Photonics}\
  }\textbf {\bibinfo {volume} {6}},\ \bibinfo {pages} {488} (\bibinfo {year}
  {2012})}\BibitemShut {NoStop}%
\bibitem [{\citenamefont {Willner}\ \emph {et~al.}(2015)\citenamefont
  {Willner}, \citenamefont {Huang}, \citenamefont {Yan}, \citenamefont {Ren},
  \citenamefont {Ahmed}, \citenamefont {Xie}, \citenamefont {Bao},
  \citenamefont {Li}, \citenamefont {Cao}, \citenamefont {Zhao} \emph
  {et~al.}}]{willner2015optical}%
  \BibitemOpen
  \bibfield  {author} {\bibinfo {author} {\bibfnamefont {A.~E.}\ \bibnamefont
  {Willner}}, \bibinfo {author} {\bibfnamefont {H.}~\bibnamefont {Huang}},
  \bibinfo {author} {\bibfnamefont {Y.}~\bibnamefont {Yan}}, \bibinfo {author}
  {\bibfnamefont {Y.}~\bibnamefont {Ren}}, \bibinfo {author} {\bibfnamefont
  {N.}~\bibnamefont {Ahmed}}, \bibinfo {author} {\bibfnamefont
  {G.}~\bibnamefont {Xie}}, \bibinfo {author} {\bibfnamefont {C.}~\bibnamefont
  {Bao}}, \bibinfo {author} {\bibfnamefont {L.}~\bibnamefont {Li}}, \bibinfo
  {author} {\bibfnamefont {Y.}~\bibnamefont {Cao}}, \bibinfo {author}
  {\bibfnamefont {Z.}~\bibnamefont {Zhao}},  \emph {et~al.},\ }\href@noop {}
  {\bibfield  {journal} {\bibinfo  {journal} {Advances in Optics and
  Photonics}\ }\textbf {\bibinfo {volume} {7}},\ \bibinfo {pages} {66}
  (\bibinfo {year} {2015})}\BibitemShut {NoStop}%
\bibitem [{\citenamefont {Mair}\ \emph {et~al.}(2001)\citenamefont {Mair},
  \citenamefont {Vaziri}, \citenamefont {Weihs},\ and\ \citenamefont
  {Zeilinger}}]{mair2001entanglement}%
  \BibitemOpen
  \bibfield  {author} {\bibinfo {author} {\bibfnamefont {A.}~\bibnamefont
  {Mair}}, \bibinfo {author} {\bibfnamefont {A.}~\bibnamefont {Vaziri}},
  \bibinfo {author} {\bibfnamefont {G.}~\bibnamefont {Weihs}}, \ and\ \bibinfo
  {author} {\bibfnamefont {A.}~\bibnamefont {Zeilinger}},\ }\href@noop {}
  {\bibfield  {journal} {\bibinfo  {journal} {Nature}\ }\textbf {\bibinfo
  {volume} {412}},\ \bibinfo {pages} {313} (\bibinfo {year}
  {2001})}\BibitemShut {NoStop}%
\bibitem [{\citenamefont {Krenn}\ \emph {et~al.}(2014)\citenamefont {Krenn},
  \citenamefont {Huber}, \citenamefont {Fickler}, \citenamefont {Lapkiewicz},
  \citenamefont {Ramelow},\ and\ \citenamefont
  {Zeilinger}}]{krenn2014generation}%
  \BibitemOpen
  \bibfield  {author} {\bibinfo {author} {\bibfnamefont {M.}~\bibnamefont
  {Krenn}}, \bibinfo {author} {\bibfnamefont {M.}~\bibnamefont {Huber}},
  \bibinfo {author} {\bibfnamefont {R.}~\bibnamefont {Fickler}}, \bibinfo
  {author} {\bibfnamefont {R.}~\bibnamefont {Lapkiewicz}}, \bibinfo {author}
  {\bibfnamefont {S.}~\bibnamefont {Ramelow}}, \ and\ \bibinfo {author}
  {\bibfnamefont {A.}~\bibnamefont {Zeilinger}},\ }\href@noop {} {\bibfield
  {journal} {\bibinfo  {journal} {Proceedings of the National Academy of
  Sciences}\ }\textbf {\bibinfo {volume} {111}},\ \bibinfo {pages} {6243}
  (\bibinfo {year} {2014})}\BibitemShut {NoStop}%
\bibitem [{\citenamefont {Mirhosseini}\ \emph {et~al.}(2015)\citenamefont
  {Mirhosseini}, \citenamefont {Maga{\~n}a-Loaiza}, \citenamefont {O'Sullivan},
  \citenamefont {Rodenburg}, \citenamefont {Malik}, \citenamefont {Lavery},
  \citenamefont {Padgett}, \citenamefont {Gauthier},\ and\ \citenamefont
  {Boyd}}]{mirhosseini2015high}%
  \BibitemOpen
  \bibfield  {author} {\bibinfo {author} {\bibfnamefont {M.}~\bibnamefont
  {Mirhosseini}}, \bibinfo {author} {\bibfnamefont {O.~S.}\ \bibnamefont
  {Maga{\~n}a-Loaiza}}, \bibinfo {author} {\bibfnamefont {M.~N.}\ \bibnamefont
  {O'Sullivan}}, \bibinfo {author} {\bibfnamefont {B.}~\bibnamefont
  {Rodenburg}}, \bibinfo {author} {\bibfnamefont {M.}~\bibnamefont {Malik}},
  \bibinfo {author} {\bibfnamefont {M.~P.}\ \bibnamefont {Lavery}}, \bibinfo
  {author} {\bibfnamefont {M.~J.}\ \bibnamefont {Padgett}}, \bibinfo {author}
  {\bibfnamefont {D.~J.}\ \bibnamefont {Gauthier}}, \ and\ \bibinfo {author}
  {\bibfnamefont {R.~W.}\ \bibnamefont {Boyd}},\ }\href@noop {} {\bibfield
  {journal} {\bibinfo  {journal} {New Journal of Physics}\ }\textbf {\bibinfo
  {volume} {17}},\ \bibinfo {pages} {033033} (\bibinfo {year}
  {2015})}\BibitemShut {NoStop}%
\bibitem [{\citenamefont {Cardano}\ \emph {et~al.}(2015)\citenamefont
  {Cardano}, \citenamefont {Massa}, \citenamefont {Qassim}, \citenamefont
  {Karimi}, \citenamefont {Slussarenko}, \citenamefont {Paparo}, \citenamefont
  {de~Lisio}, \citenamefont {Sciarrino}, \citenamefont {Santamato},
  \citenamefont {Boyd} \emph {et~al.}}]{cardano2015quantum}%
  \BibitemOpen
  \bibfield  {author} {\bibinfo {author} {\bibfnamefont {F.}~\bibnamefont
  {Cardano}}, \bibinfo {author} {\bibfnamefont {F.}~\bibnamefont {Massa}},
  \bibinfo {author} {\bibfnamefont {H.}~\bibnamefont {Qassim}}, \bibinfo
  {author} {\bibfnamefont {E.}~\bibnamefont {Karimi}}, \bibinfo {author}
  {\bibfnamefont {S.}~\bibnamefont {Slussarenko}}, \bibinfo {author}
  {\bibfnamefont {D.}~\bibnamefont {Paparo}}, \bibinfo {author} {\bibfnamefont
  {C.}~\bibnamefont {de~Lisio}}, \bibinfo {author} {\bibfnamefont
  {F.}~\bibnamefont {Sciarrino}}, \bibinfo {author} {\bibfnamefont
  {E.}~\bibnamefont {Santamato}}, \bibinfo {author} {\bibfnamefont {R.~W.}\
  \bibnamefont {Boyd}},  \emph {et~al.},\ }\href@noop {} {\bibfield  {journal}
  {\bibinfo  {journal} {Science advances}\ }\textbf {\bibinfo {volume} {1}},\
  \bibinfo {pages} {e1500087} (\bibinfo {year} {2015})}\BibitemShut {NoStop}%
\bibitem [{\citenamefont {Malik}\ \emph {et~al.}(2016)\citenamefont {Malik},
  \citenamefont {Erhard}, \citenamefont {Huber}, \citenamefont {Krenn},
  \citenamefont {Fickler},\ and\ \citenamefont {Zeilinger}}]{malik2016multi}%
  \BibitemOpen
  \bibfield  {author} {\bibinfo {author} {\bibfnamefont {M.}~\bibnamefont
  {Malik}}, \bibinfo {author} {\bibfnamefont {M.}~\bibnamefont {Erhard}},
  \bibinfo {author} {\bibfnamefont {M.}~\bibnamefont {Huber}}, \bibinfo
  {author} {\bibfnamefont {M.}~\bibnamefont {Krenn}}, \bibinfo {author}
  {\bibfnamefont {R.}~\bibnamefont {Fickler}}, \ and\ \bibinfo {author}
  {\bibfnamefont {A.}~\bibnamefont {Zeilinger}},\ }\href@noop {} {\bibfield
  {journal} {\bibinfo  {journal} {Nature Photonics}\ }\textbf {\bibinfo
  {volume} {10}},\ \bibinfo {pages} {248} (\bibinfo {year} {2016})}\BibitemShut
  {NoStop}%
\bibitem [{\citenamefont {Forbes}\ \emph {et~al.}(2016)\citenamefont {Forbes},
  \citenamefont {Dudley},\ and\ \citenamefont {McLaren}}]{forbes2016creation}%
  \BibitemOpen
  \bibfield  {author} {\bibinfo {author} {\bibfnamefont {A.}~\bibnamefont
  {Forbes}}, \bibinfo {author} {\bibfnamefont {A.}~\bibnamefont {Dudley}}, \
  and\ \bibinfo {author} {\bibfnamefont {M.}~\bibnamefont {McLaren}},\
  }\href@noop {} {\bibfield  {journal} {\bibinfo  {journal} {Advances in Optics
  and Photonics}\ }\textbf {\bibinfo {volume} {8}},\ \bibinfo {pages} {200}
  (\bibinfo {year} {2016})}\BibitemShut {NoStop}%
\bibitem [{\citenamefont {Morizur}\ \emph {et~al.}(2010)\citenamefont
  {Morizur}, \citenamefont {Nicholls}, \citenamefont {Jian}, \citenamefont
  {Armstrong}, \citenamefont {Treps}, \citenamefont {Hage}, \citenamefont
  {Hsu}, \citenamefont {Bowen}, \citenamefont {Janousek},\ and\ \citenamefont
  {Bachor}}]{morizur2010programmable}%
  \BibitemOpen
  \bibfield  {author} {\bibinfo {author} {\bibfnamefont {J.-F.}\ \bibnamefont
  {Morizur}}, \bibinfo {author} {\bibfnamefont {L.}~\bibnamefont {Nicholls}},
  \bibinfo {author} {\bibfnamefont {P.}~\bibnamefont {Jian}}, \bibinfo {author}
  {\bibfnamefont {S.}~\bibnamefont {Armstrong}}, \bibinfo {author}
  {\bibfnamefont {N.}~\bibnamefont {Treps}}, \bibinfo {author} {\bibfnamefont
  {B.}~\bibnamefont {Hage}}, \bibinfo {author} {\bibfnamefont {M.}~\bibnamefont
  {Hsu}}, \bibinfo {author} {\bibfnamefont {W.}~\bibnamefont {Bowen}}, \bibinfo
  {author} {\bibfnamefont {J.}~\bibnamefont {Janousek}}, \ and\ \bibinfo
  {author} {\bibfnamefont {H.-A.}\ \bibnamefont {Bachor}},\ }\href@noop {}
  {\bibfield  {journal} {\bibinfo  {journal} {JOSA A}\ }\textbf {\bibinfo
  {volume} {27}},\ \bibinfo {pages} {2524} (\bibinfo {year}
  {2010})}\BibitemShut {NoStop}%
\bibitem [{\citenamefont {Leach}\ \emph {et~al.}(2002)\citenamefont {Leach},
  \citenamefont {Padgett}, \citenamefont {Barnett}, \citenamefont
  {Franke-Arnold},\ and\ \citenamefont {Courtial}}]{leach2002measuring}%
  \BibitemOpen
  \bibfield  {author} {\bibinfo {author} {\bibfnamefont {J.}~\bibnamefont
  {Leach}}, \bibinfo {author} {\bibfnamefont {M.~J.}\ \bibnamefont {Padgett}},
  \bibinfo {author} {\bibfnamefont {S.~M.}\ \bibnamefont {Barnett}}, \bibinfo
  {author} {\bibfnamefont {S.}~\bibnamefont {Franke-Arnold}}, \ and\ \bibinfo
  {author} {\bibfnamefont {J.}~\bibnamefont {Courtial}},\ }\href@noop {}
  {\bibfield  {journal} {\bibinfo  {journal} {Physical review letters}\
  }\textbf {\bibinfo {volume} {88}},\ \bibinfo {pages} {257901} (\bibinfo
  {year} {2002})}\BibitemShut {NoStop}%
\bibitem [{\citenamefont {Schlederer}\ \emph {et~al.}(2016)\citenamefont
  {Schlederer}, \citenamefont {Krenn}, \citenamefont {Fickler}, \citenamefont
  {Malik},\ and\ \citenamefont {Zeilinger}}]{schlederer2016cyclic}%
  \BibitemOpen
  \bibfield  {author} {\bibinfo {author} {\bibfnamefont {F.}~\bibnamefont
  {Schlederer}}, \bibinfo {author} {\bibfnamefont {M.}~\bibnamefont {Krenn}},
  \bibinfo {author} {\bibfnamefont {R.}~\bibnamefont {Fickler}}, \bibinfo
  {author} {\bibfnamefont {M.}~\bibnamefont {Malik}}, \ and\ \bibinfo {author}
  {\bibfnamefont {A.}~\bibnamefont {Zeilinger}},\ }\href@noop {} {\bibfield
  {journal} {\bibinfo  {journal} {New Journal of Physics}\ }\textbf {\bibinfo
  {volume} {18}},\ \bibinfo {pages} {043019} (\bibinfo {year}
  {2016})}\BibitemShut {NoStop}%
\bibitem [{\citenamefont {Fickler}\ \emph {et~al.}(2014)\citenamefont
  {Fickler}, \citenamefont {Lapkiewicz}, \citenamefont {Huber}, \citenamefont
  {Lavery}, \citenamefont {Padgett},\ and\ \citenamefont
  {Zeilinger}}]{fickler2014interface}%
  \BibitemOpen
  \bibfield  {author} {\bibinfo {author} {\bibfnamefont {R.}~\bibnamefont
  {Fickler}}, \bibinfo {author} {\bibfnamefont {R.}~\bibnamefont {Lapkiewicz}},
  \bibinfo {author} {\bibfnamefont {M.}~\bibnamefont {Huber}}, \bibinfo
  {author} {\bibfnamefont {M.~P.}\ \bibnamefont {Lavery}}, \bibinfo {author}
  {\bibfnamefont {M.~J.}\ \bibnamefont {Padgett}}, \ and\ \bibinfo {author}
  {\bibfnamefont {A.}~\bibnamefont {Zeilinger}},\ }\href@noop {} {\bibfield
  {journal} {\bibinfo  {journal} {Nature communications}\ }\textbf {\bibinfo
  {volume} {5}} (\bibinfo {year} {2014})}\BibitemShut {NoStop}%
\bibitem [{\citenamefont {Malik}\ \emph {et~al.}(2014)\citenamefont {Malik},
  \citenamefont {Mirhosseini}, \citenamefont {Lavery}, \citenamefont {Leach},
  \citenamefont {Padgett},\ and\ \citenamefont {Boyd}}]{malik2014direct}%
  \BibitemOpen
  \bibfield  {author} {\bibinfo {author} {\bibfnamefont {M.}~\bibnamefont
  {Malik}}, \bibinfo {author} {\bibfnamefont {M.}~\bibnamefont {Mirhosseini}},
  \bibinfo {author} {\bibfnamefont {M.~P.}\ \bibnamefont {Lavery}}, \bibinfo
  {author} {\bibfnamefont {J.}~\bibnamefont {Leach}}, \bibinfo {author}
  {\bibfnamefont {M.~J.}\ \bibnamefont {Padgett}}, \ and\ \bibinfo {author}
  {\bibfnamefont {R.~W.}\ \bibnamefont {Boyd}},\ }\href@noop {} {\bibfield
  {journal} {\bibinfo  {journal} {Nature communications}\ }\textbf {\bibinfo
  {volume} {5}} (\bibinfo {year} {2014})}\BibitemShut {NoStop}%
\bibitem [{\citenamefont {Labroille}\ \emph {et~al.}(2014)\citenamefont
  {Labroille}, \citenamefont {Denolle}, \citenamefont {Jian}, \citenamefont
  {Genevaux}, \citenamefont {Treps},\ and\ \citenamefont
  {Morizur}}]{labroille2014efficient}%
  \BibitemOpen
  \bibfield  {author} {\bibinfo {author} {\bibfnamefont {G.}~\bibnamefont
  {Labroille}}, \bibinfo {author} {\bibfnamefont {B.}~\bibnamefont {Denolle}},
  \bibinfo {author} {\bibfnamefont {P.}~\bibnamefont {Jian}}, \bibinfo {author}
  {\bibfnamefont {P.}~\bibnamefont {Genevaux}}, \bibinfo {author}
  {\bibfnamefont {N.}~\bibnamefont {Treps}}, \ and\ \bibinfo {author}
  {\bibfnamefont {J.-F.}\ \bibnamefont {Morizur}},\ }\href@noop {} {\bibfield
  {journal} {\bibinfo  {journal} {Optics express}\ }\textbf {\bibinfo {volume}
  {22}},\ \bibinfo {pages} {15599} (\bibinfo {year} {2014})}\BibitemShut
  {NoStop}%
\bibitem [{\citenamefont {Berkhout}\ \emph {et~al.}(2010)\citenamefont
  {Berkhout}, \citenamefont {Lavery}, \citenamefont {Courtial}, \citenamefont
  {Beijersbergen},\ and\ \citenamefont {Padgett}}]{berkhout2010efficient}%
  \BibitemOpen
  \bibfield  {author} {\bibinfo {author} {\bibfnamefont {G.~C.~G.}\
  \bibnamefont {Berkhout}}, \bibinfo {author} {\bibfnamefont {M.~P.~J.}\
  \bibnamefont {Lavery}}, \bibinfo {author} {\bibfnamefont {J.}~\bibnamefont
  {Courtial}}, \bibinfo {author} {\bibfnamefont {M.~W.}\ \bibnamefont
  {Beijersbergen}}, \ and\ \bibinfo {author} {\bibfnamefont {M.~J.}\
  \bibnamefont {Padgett}},\ }\href@noop {} {\bibfield  {journal} {\bibinfo
  {journal} {Physical Review Letters}\ }\textbf {\bibinfo {volume} {105}},\
  \bibinfo {pages} {153601} (\bibinfo {year} {2010})}\BibitemShut {NoStop}%
\bibitem [{\citenamefont {Mirhosseini}\ \emph {et~al.}(2013)\citenamefont
  {Mirhosseini}, \citenamefont {Malik}, \citenamefont {Shi},\ and\
  \citenamefont {Boyd}}]{mirhosseini2013efficient}%
  \BibitemOpen
  \bibfield  {author} {\bibinfo {author} {\bibfnamefont {M.}~\bibnamefont
  {Mirhosseini}}, \bibinfo {author} {\bibfnamefont {M.}~\bibnamefont {Malik}},
  \bibinfo {author} {\bibfnamefont {Z.}~\bibnamefont {Shi}}, \ and\ \bibinfo
  {author} {\bibfnamefont {R.~W.}\ \bibnamefont {Boyd}},\ }\href@noop {}
  {\bibfield  {journal} {\bibinfo  {journal} {Nature communications}\ }\textbf
  {\bibinfo {volume} {4}} (\bibinfo {year} {2013})}\BibitemShut {NoStop}%
\bibitem [{\citenamefont {Mosk}\ \emph {et~al.}(2012)\citenamefont {Mosk},
  \citenamefont {Lagendijk}, \citenamefont {Lerosey},\ and\ \citenamefont
  {Fink}}]{mosk2012controlling}%
  \BibitemOpen
  \bibfield  {author} {\bibinfo {author} {\bibfnamefont {A.~P.}\ \bibnamefont
  {Mosk}}, \bibinfo {author} {\bibfnamefont {A.}~\bibnamefont {Lagendijk}},
  \bibinfo {author} {\bibfnamefont {G.}~\bibnamefont {Lerosey}}, \ and\
  \bibinfo {author} {\bibfnamefont {M.}~\bibnamefont {Fink}},\ }\href@noop {}
  {\bibfield  {journal} {\bibinfo  {journal} {Nature photonics}\ }\textbf
  {\bibinfo {volume} {6}},\ \bibinfo {pages} {283} (\bibinfo {year}
  {2012})}\BibitemShut {NoStop}%
\bibitem [{\citenamefont {Vellekoop}\ \emph {et~al.}(2010)\citenamefont
  {Vellekoop}, \citenamefont {Lagendijk},\ and\ \citenamefont
  {Mosk}}]{vellekoop2010exploiting}%
  \BibitemOpen
  \bibfield  {author} {\bibinfo {author} {\bibfnamefont {I.}~\bibnamefont
  {Vellekoop}}, \bibinfo {author} {\bibfnamefont {A.}~\bibnamefont
  {Lagendijk}}, \ and\ \bibinfo {author} {\bibfnamefont {A.}~\bibnamefont
  {Mosk}},\ }\href@noop {} {\bibfield  {journal} {\bibinfo  {journal} {Nature
  photonics}\ }\textbf {\bibinfo {volume} {4}},\ \bibinfo {pages} {320}
  (\bibinfo {year} {2010})}\BibitemShut {NoStop}%
\bibitem [{\citenamefont {Bertolotti}\ \emph {et~al.}(2012)\citenamefont
  {Bertolotti}, \citenamefont {van Putten}, \citenamefont {Blum}, \citenamefont
  {Lagendijk}, \citenamefont {Vos},\ and\ \citenamefont
  {Mosk}}]{bertolotti2012non}%
  \BibitemOpen
  \bibfield  {author} {\bibinfo {author} {\bibfnamefont {J.}~\bibnamefont
  {Bertolotti}}, \bibinfo {author} {\bibfnamefont {E.~G.}\ \bibnamefont {van
  Putten}}, \bibinfo {author} {\bibfnamefont {C.}~\bibnamefont {Blum}},
  \bibinfo {author} {\bibfnamefont {A.}~\bibnamefont {Lagendijk}}, \bibinfo
  {author} {\bibfnamefont {W.~L.}\ \bibnamefont {Vos}}, \ and\ \bibinfo
  {author} {\bibfnamefont {A.~P.}\ \bibnamefont {Mosk}},\ }\href@noop {}
  {\bibfield  {journal} {\bibinfo  {journal} {Nature}\ }\textbf {\bibinfo
  {volume} {491}},\ \bibinfo {pages} {232} (\bibinfo {year}
  {2012})}\BibitemShut {NoStop}%
\bibitem [{\citenamefont {Katz}\ \emph {et~al.}(2012)\citenamefont {Katz},
  \citenamefont {Small},\ and\ \citenamefont {Silberberg}}]{katz2012looking}%
  \BibitemOpen
  \bibfield  {author} {\bibinfo {author} {\bibfnamefont {O.}~\bibnamefont
  {Katz}}, \bibinfo {author} {\bibfnamefont {E.}~\bibnamefont {Small}}, \ and\
  \bibinfo {author} {\bibfnamefont {Y.}~\bibnamefont {Silberberg}},\
  }\href@noop {} {\bibfield  {journal} {\bibinfo  {journal} {Nature photonics}\
  }\textbf {\bibinfo {volume} {6}},\ \bibinfo {pages} {549} (\bibinfo {year}
  {2012})}\BibitemShut {NoStop}%
\bibitem [{\citenamefont {Katz}\ \emph {et~al.}(2011)\citenamefont {Katz},
  \citenamefont {Small}, \citenamefont {Bromberg},\ and\ \citenamefont
  {Silberberg}}]{katz2011focusing}%
  \BibitemOpen
  \bibfield  {author} {\bibinfo {author} {\bibfnamefont {O.}~\bibnamefont
  {Katz}}, \bibinfo {author} {\bibfnamefont {E.}~\bibnamefont {Small}},
  \bibinfo {author} {\bibfnamefont {Y.}~\bibnamefont {Bromberg}}, \ and\
  \bibinfo {author} {\bibfnamefont {Y.}~\bibnamefont {Silberberg}},\
  }\href@noop {} {\bibfield  {journal} {\bibinfo  {journal} {Nature photonics}\
  }\textbf {\bibinfo {volume} {5}},\ \bibinfo {pages} {372} (\bibinfo {year}
  {2011})}\BibitemShut {NoStop}%
\bibitem [{\citenamefont {Huisman}\ \emph {et~al.}(2015)\citenamefont
  {Huisman}, \citenamefont {Huisman}, \citenamefont {Wolterink}, \citenamefont
  {Mosk},\ and\ \citenamefont {Pinkse}}]{huisman2015programmable}%
  \BibitemOpen
  \bibfield  {author} {\bibinfo {author} {\bibfnamefont {S.~R.}\ \bibnamefont
  {Huisman}}, \bibinfo {author} {\bibfnamefont {T.~J.}\ \bibnamefont
  {Huisman}}, \bibinfo {author} {\bibfnamefont {T.~A.}\ \bibnamefont
  {Wolterink}}, \bibinfo {author} {\bibfnamefont {A.~P.}\ \bibnamefont {Mosk}},
  \ and\ \bibinfo {author} {\bibfnamefont {P.~W.}\ \bibnamefont {Pinkse}},\
  }\href@noop {} {\bibfield  {journal} {\bibinfo  {journal} {Optics express}\
  }\textbf {\bibinfo {volume} {23}},\ \bibinfo {pages} {3102} (\bibinfo {year}
  {2015})}\BibitemShut {NoStop}%
\bibitem [{\citenamefont {Defienne}\ \emph {et~al.}(2016)\citenamefont
  {Defienne}, \citenamefont {Barbieri}, \citenamefont {Walmsley}, \citenamefont
  {Smith},\ and\ \citenamefont {Gigan}}]{defienne2016two}%
  \BibitemOpen
  \bibfield  {author} {\bibinfo {author} {\bibfnamefont {H.}~\bibnamefont
  {Defienne}}, \bibinfo {author} {\bibfnamefont {M.}~\bibnamefont {Barbieri}},
  \bibinfo {author} {\bibfnamefont {I.~A.}\ \bibnamefont {Walmsley}}, \bibinfo
  {author} {\bibfnamefont {B.~J.}\ \bibnamefont {Smith}}, \ and\ \bibinfo
  {author} {\bibfnamefont {S.}~\bibnamefont {Gigan}},\ }\href@noop {}
  {\bibfield  {journal} {\bibinfo  {journal} {Science advances}\ }\textbf
  {\bibinfo {volume} {2}},\ \bibinfo {pages} {e1501054} (\bibinfo {year}
  {2016})}\BibitemShut {NoStop}%
\bibitem [{\citenamefont {Bolduc}\ \emph {et~al.}(2013)\citenamefont {Bolduc},
  \citenamefont {Bent}, \citenamefont {Santamato}, \citenamefont {Karimi},\
  and\ \citenamefont {Boyd}}]{bolduc2013exact}%
  \BibitemOpen
  \bibfield  {author} {\bibinfo {author} {\bibfnamefont {E.}~\bibnamefont
  {Bolduc}}, \bibinfo {author} {\bibfnamefont {N.}~\bibnamefont {Bent}},
  \bibinfo {author} {\bibfnamefont {E.}~\bibnamefont {Santamato}}, \bibinfo
  {author} {\bibfnamefont {E.}~\bibnamefont {Karimi}}, \ and\ \bibinfo {author}
  {\bibfnamefont {R.~W.}\ \bibnamefont {Boyd}},\ }\href@noop {} {\bibfield
  {journal} {\bibinfo  {journal} {Optics Letters}\ }\textbf {\bibinfo {volume}
  {38}},\ \bibinfo {pages} {3546} (\bibinfo {year} {2013})}\BibitemShut
  {NoStop}%
\bibitem [{\citenamefont {Kop}\ \emph {et~al.}(1997)\citenamefont {Kop},
  \citenamefont {de~Vries}, \citenamefont {Sprik},\ and\ \citenamefont
  {Lagendijk}}]{kop1997observation}%
  \BibitemOpen
  \bibfield  {author} {\bibinfo {author} {\bibfnamefont {R.~H.~J.}\
  \bibnamefont {Kop}}, \bibinfo {author} {\bibfnamefont {P.}~\bibnamefont
  {de~Vries}}, \bibinfo {author} {\bibfnamefont {R.}~\bibnamefont {Sprik}}, \
  and\ \bibinfo {author} {\bibfnamefont {A.}~\bibnamefont {Lagendijk}},\
  }\href@noop {} {\bibfield  {journal} {\bibinfo  {journal} {Physical Review
  Letters}\ }\textbf {\bibinfo {volume} {79}},\ \bibinfo {pages} {4369}
  (\bibinfo {year} {1997})}\BibitemShut {NoStop}%
\bibitem [{\citenamefont {Conkey}\ \emph {et~al.}(2012)\citenamefont {Conkey},
  \citenamefont {Brown}, \citenamefont {Caravaca-Aguirre},\ and\ \citenamefont
  {Piestun}}]{conkey2012genetic}%
  \BibitemOpen
  \bibfield  {author} {\bibinfo {author} {\bibfnamefont {D.~B.}\ \bibnamefont
  {Conkey}}, \bibinfo {author} {\bibfnamefont {A.~N.}\ \bibnamefont {Brown}},
  \bibinfo {author} {\bibfnamefont {A.~M.}\ \bibnamefont {Caravaca-Aguirre}}, \
  and\ \bibinfo {author} {\bibfnamefont {R.}~\bibnamefont {Piestun}},\
  }\href@noop {} {\bibfield  {journal} {\bibinfo  {journal} {Optics express}\
  }\textbf {\bibinfo {volume} {20}},\ \bibinfo {pages} {4840} (\bibinfo {year}
  {2012})}\BibitemShut {NoStop}%
\bibitem [{\citenamefont {Zhao}\ \emph {et~al.}(2015)\citenamefont {Zhao},
  \citenamefont {Li}, \citenamefont {Li},\ and\ \citenamefont
  {Kahn}}]{zhao2015capacity}%
  \BibitemOpen
  \bibfield  {author} {\bibinfo {author} {\bibfnamefont {N.}~\bibnamefont
  {Zhao}}, \bibinfo {author} {\bibfnamefont {X.}~\bibnamefont {Li}}, \bibinfo
  {author} {\bibfnamefont {G.}~\bibnamefont {Li}}, \ and\ \bibinfo {author}
  {\bibfnamefont {J.~M.}\ \bibnamefont {Kahn}},\ }\href@noop {} {\bibfield
  {journal} {\bibinfo  {journal} {Nature Photonics}\ }\textbf {\bibinfo
  {volume} {9}},\ \bibinfo {pages} {822} (\bibinfo {year} {2015})}\BibitemShut
  {NoStop}%
\bibitem [{\citenamefont {Kahn}\ \emph {et~al.}(2016)\citenamefont {Kahn},
  \citenamefont {Li}, \citenamefont {Li},\ and\ \citenamefont
  {Zhao}}]{kahn2016twist}%
  \BibitemOpen
  \bibfield  {author} {\bibinfo {author} {\bibfnamefont {J.~M.}\ \bibnamefont
  {Kahn}}, \bibinfo {author} {\bibfnamefont {G.}~\bibnamefont {Li}}, \bibinfo
  {author} {\bibfnamefont {X.}~\bibnamefont {Li}}, \ and\ \bibinfo {author}
  {\bibfnamefont {N.}~\bibnamefont {Zhao}},\ }in\ \href@noop {} {\emph
  {\bibinfo {booktitle} {Signal Processing in Photonic Communications}}}\
  (\bibinfo {organization} {Optical Society of America},\ \bibinfo {year}
  {2016})\ pp.\ \bibinfo {pages} {SpM4E--1}\BibitemShut {NoStop}%
\end{thebibliography}%

\end{document}